\documentclass[10pt,journal,compsoc]{IEEEtran}
\usepackage{amsmath,amsfonts}
\usepackage{algorithmic}
\usepackage{algorithm}
\usepackage{array}
\usepackage[caption=false,font=normalsize,labelfont=sf,textfont=sf]{subfig}
\usepackage{textcomp}
\usepackage{stfloats}
\usepackage{url}
\usepackage{verbatim}
\usepackage{graphicx}
\usepackage{cite}
\usepackage{multirow}
\usepackage{color}
\usepackage{hyperref}

\begin{document}

\title{\textsc{PoTable}: Towards Systematic Thinking via Plan-then-Execute Stage Reasoning on Tables}

\author{Qingyang Mao, Qi Liu,~\IEEEmembership{Member,~IEEE}, Zhi Li, Mingyue Cheng, Zheng Zhang, Rui Li
        % <-this % stops a space
\IEEEcompsocitemizethanks{
\IEEEcompsocthanksitem Qingyang Mao, Qi Liu (corresponding author), Mingyue Cheng, Zheng Zhang, Rui Li are with State Key Laboratory of Cognitive Intelligence, University of Science and Technology of China, Hefei, Anhui, China. E-mail: maoqy0503@mail.ustc.edu.cn, qiliuql@ustc.edu.cn, mycheng@ustc.edu.cn, zhangzheng@mail.ustc.edu.cn, ruili2000@mail.ustc.edu.cn. 
Zhi Li is with Shenzhen International Graduate School, Tsinghua University, Shenzhen, Guangdong, China. Email: zhilizl@sz.tsinghua.edu.cn. 
}
}

\markboth{IEEE Transactions on Knowledge and Data Engineering 2026}%
{Shell \MakeLowercase{\textit{et al.}}: A Sample Article Using IEEEtran.cls for IEEE Journals}


\IEEEtitleabstractindextext{
\begin{abstract}
    In recent years, table reasoning has garnered substantial research interest, particularly regarding its integration with Large Language Models (LLMs), which have revolutionized natural language applications. 
    Existing LLM-based studies typically achieve step-by-step thinking for table reasoning guided by task semantics. 
    While these approaches emphasize autonomous exploration and enhance fine-grained table understanding, they often overlook systematic thinking in the reasoning process. 
    This oversight can lead to omitted steps, disorganized logic and misleading results, especially in complex scenarios. 
    In this paper, we propose \textsc{PoTable}, a novel stage-oriented plan-then-execute approach that incorporates systematic thinking into table reasoning. 
    Specifically, \textsc{PoTable} involves several distinct analytical stages with clear objectives to provide adequate guidance. 
    To accomplish stage-specific goals, \textsc{PoTable} employs a plan-then-execute mechanism: it first plans the operation chain based on the stage objective, and then executes operations sequentially through code generation, real-time running and feedback processing. 
    Consequently, \textsc{PoTable} produces reliable table reasoning results with highly accurate, step-wise commented and completely executable programs. 
    It mirrors the workflow of a professional data analyst, offering advantages in both accuracy and explainability. 
    Finally, we conduct extensive experiments on four datasets from the WikiTQ and TabFact benchmarks, where the results demonstrate the effectiveness, efficiency and explainability of \textsc{PoTable}. 
    Our code is available at: \href{https://github.com/Double680/PoTable}{https://github.com/Double680/PoTable}.
\end{abstract}

}

\maketitle

\section{Introduction}

\begin{figure*}
\begin{center}
    \includegraphics[width=520pt]{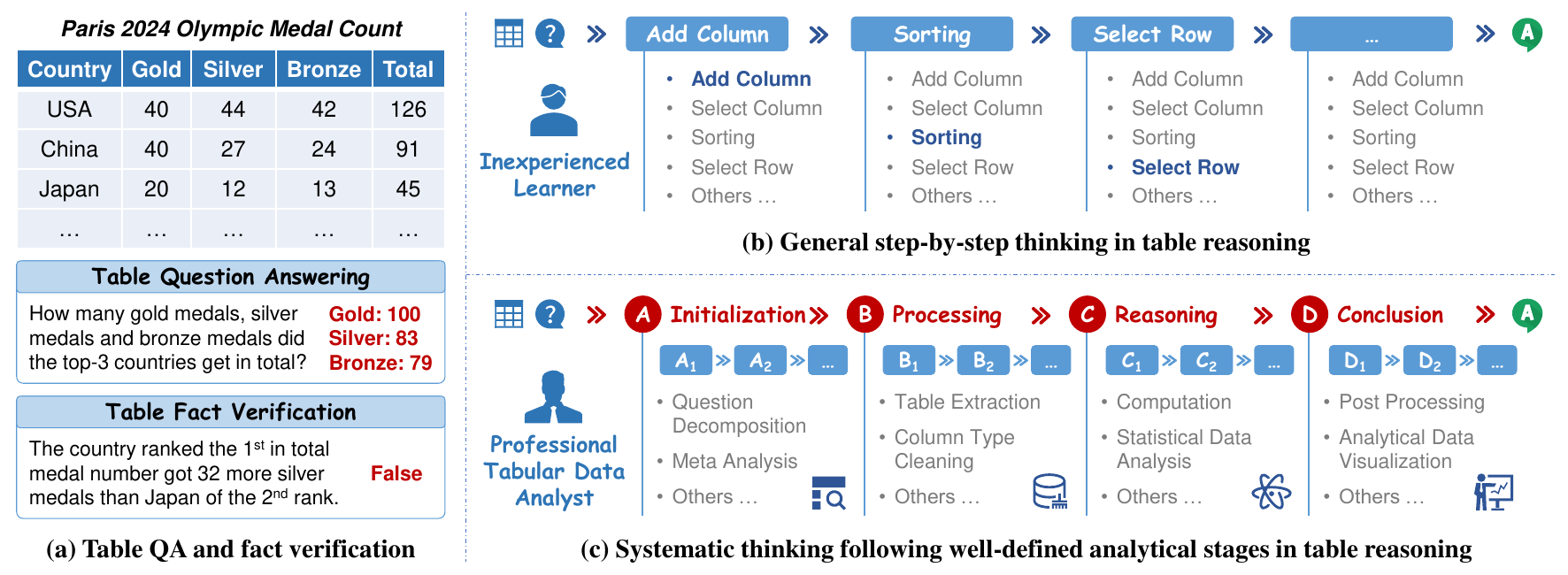}
\end{center}
\caption{Illustrations of (a) two table reasoning tasks, (b) general step-by-step thinking in typical LLM-based table reasoning studies, and (c) systematic thinking following well-defined analytical stages to reason like a professional tabular data analyst. }
\label{fig:introduction}
\end{figure*}

\IEEEPARstart{T}{ables} serve as a critical medium for presenting structured information across various domains (\textit{e.g.}, scientific literature \cite{PubTables-1M, SciTab}, medical reports \cite{emrQA, RJUA-MedDQA}, financial statements \cite{TAT-QA, ConvFinQA}). 
Driven by the widespread utility of tabular data and rapid advancements in artificial intelligence, there is a growing demand for automated table reasoning \cite{TableSum, Dater, Chain-of-Table, TableMind, STaR}, which has garnered substantial interest from both academia and industry. 
Extensive studies focus on two prominent reasoning tasks: \textit{table question answering} (QA) \cite{WikiTQ, HiTab} and \textit{table fact verification} \cite{TabFact, FEVEROUS}, as illustrated in Fig.~\ref{fig:introduction}(a). 
These tasks integrate multiple processes (\textit{e.g.}, structured information retrieval, natural language understanding, numerical computation), demonstrating multifaceted challenges. 

In recent years, the evolution of Large Language Models (LLMs) \cite{LLM-IR, LLM-Survey, LLM-Eval, KG-LLM} has introduced a novel training-free prompting paradigm, revolutionizing methods and applications in the field of table reasoning \cite{LLM-Table-Survey, LLM-Table-Tutorial, NLTableQV, TableMiningSurvey, TableMind++} in three key aspects.
First, the text generation and multi-turn interaction capabilities of LLMs are well-suited for reasoning tasks involving multiple intermediate steps. 
Second, through extensive pre-training, LLMs exhibit advanced text comprehension and logical thinking abilities, demonstrating great potential for table understanding and analysis. 
Third, compared to conventional representation learning techniques, LLM-based methods provide enhanced explainability by generating detailed textual evidence to support decision-making. 
To fully leverage these reasoning capabilities, most studies focus on breaking down the entire task into smaller, executable sub-steps for sequential completion. 
This approach draws inspiration from \textit{chain-of-thought} prompting \cite{CoT} that encourages step-by-step thinking. 
Specifically, some approaches deploy explicit table decomposition to mitigate performance degradation caused by irrelevant data \cite{Dater, TabSQLify}. 
Furthermore, other methods dynamically plan the next operation from a pre-defined candidate pool \cite{Chain-of-Table} or via free exploration \cite{ReAcTable} to complete the reasoning task autonomously. 
These studies prioritize task objectives and generate satisfactory intermediate results, thereby achieving promising performance.

However, existing studies primarily emphasize LLM-based autonomous exploration guided solely by task semantics. 
They lack a systematic thinking approach tailored for table reasoning to provide adequate guidance, particularly in complex scenarios. 
In other words, these methods neglect a structured top-down guideline to address the complexities of the overall reasoning process.
As illustrated in Fig.~\ref{fig:introduction}(b), general step-by-step reasoning determines the subsequent operation based solely on the task question and suggested action choices. 
This approach suffers from two primary limitations. 
First, when tackling complex tasks, operation chains tend to become excessively long, increasing the risk of omitted steps or erroneous details. 
Second, even if the reasoning chain yields accurate results, the underlying logic is often disorganized, making it difficult to trace the source of errors and rendering the verification process time-consuming.
Consequently, the absence of structured top-down guidance often leads to unreliable outcomes, especially in scenarios involving complex tables and task objectives. 
Metaphorically, these methods behave more like \textit{inexperienced learners} than \textit{professional data analysts}. 
They often rely on local intuition or trial-and-error without a global perspective on the data analysis lifecycle.
In contrast, Fig.~\ref{fig:introduction}(c) shows how a proficient tabular data analyst adheres to well-defined analytical stages \cite{KDD-1, KDD-2} (\textit{e.g.}, initialization, processing, reasoning, conclusion). 
Such a stage-oriented approach enables the analyst to concentrate on stage-specific objectives, minimizing the likelihood of planning irrelevant steps. 
Furthermore, systematic thinking enforces clear boundaries within the overall logic, facilitating the comprehension and review of the entire reasoning process. 
Intuitively, \textit{incorporating human-like systematic thinking into LLM-based table reasoning holds significant promise for achieving more reliable and explainable outcomes.}

Along this line, we propose \textsc{PoTable} (\textbf{P}rogramming \textbf{o}n \textbf{Table}s), a novel framework that incorporates systematic thinking into table reasoning via a stage-oriented ``plan-then-execute'' mechanism.
Specifically, \textsc{PoTable} employs distinct analytical stages with clear objectives, providing structured guidance that complements the overall task goal. 
The framework tackles reasoning tasks by fulfilling these stage-specific goals sequentially, integrating an LLM with a real-time Python interpreter.
For each stage, the process operates in two phases: planning and execution.
In the planning phase, \textsc{PoTable} formulates an operation chain tailored to the specific stage objective.
As for the executing phase, \textsc{PoTable} generates corresponding code for each operation and executes it immediately via the Python interpreter.
When execution feedback reveals erroneous information, \textsc{PoTable} enters an iterative loop to regenerate the code until the issue is resolved.
Collectively, these mechanisms render \textsc{PoTable} a streamlined yet resilient framework for LLM-based table reasoning.

By adhering to this systematic framework, \textsc{PoTable} produces high-quality results accompanied by thoroughly commented and fully executable code.
This agentic workflow aligns closely with that of a professional data analyst, offering two primary advantages:
(i) \textbf{Accuracy}: Planning coherent operations under precise stage-specific goals reduces the complexity of individual steps. 
As the stage-specific operation chain tends to be shorter than full-task chains, the risk of omitted steps or hallucinations is minimized, leading to more reliable final outcomes.
(ii) \textbf{Explainability}: The generation of executable code with distinct stage boundaries provides structured evidence, facilitating the verification of the reasoning completeness and accuracy.
Finally, extensive experiments on four datasets from the WikiTQ and TabFact benchmarks demonstrate the superiority of \textsc{PoTable} over existing LLM-based baselines.
Notably, GPT-based \textsc{PoTable} achieves an absolute accuracy gain of over 4.3\% on standard datasets and a 3.68\% improvement on the complex evaluation set compared to the runner-up.
These comprehensive analyses confirm the strong effectiveness, efficiency, and explainability of our proposed approach.

Our main contributions can be summarized as follows:
\begin{itemize}
    \item We propose \textsc{PoTable}, a novel LLM-based table reasoning approach that integrates systematic thinking. 
    Distinct analytical stages provide critical structured guidance, enabling \textsc{PoTable} to concentrate on stage-specific goals and minimize the risk of omitted steps or misleading details.
    \item \textsc{PoTable} implements a plan-then-execute reasoning paradigm through the collaboration of an LLM and a real-time Python interpreter. 
    \textsc{PoTable} can produce highly accurate, step-wise commented, and fully executable programs, demonstrating significant advantages in terms of accuracy and explainability.  
    \item Extensive experiments on datasets from the WikiTQ and TabFact benchmarks demonstrate the promising performance of \textsc{PoTable}, showing its effectiveness, efficiency and explainability. 
\end{itemize}

\section{Related Work}

Table reasoning has gained significant attention as a research area over the past decade. 
In this field, language models and symbolic tools are widely adopted as crucial components in processes like structured information retrieval, natural language understanding and numerical computation. 
This section describes related studies on table reasoning, focusing on the utilization of language models and symbolic tools.

\subsection{Table Reasoning with Language Models}

In the field of table reasoning, language models play a significant role in processing and understanding the flattened text-form tabular data. 
Before the era of LLMs, numerous efforts were made to process tabular data with pre-trained language models. 
TaPas \cite{TaPas} is a weakly supervised table QA model based on BERT \cite{BERT}, which predicts denotations by selecting cells and aggregation operations with a joint pre-training of text segments and tables. 
TaBERT \cite{TaBERT} combines content snapshot and vertical attention within BERT to obtain joint textual and tabular representations for further understanding. 
TUTA \cite{TUTA} enhances Transformer \cite{Transformer} with structure-aware mechanisms to effectively capture spatial, hierarchical and semantic information.
TAPEX \cite{TAPEX} pre-trains BART \cite{BART} on a large synthetic SQL dataset to imitate the SQL executor that better understands tabular structure information. 
With the development of LLMs, the paradigm of table reasoning has been fundamentally transformed, especially in table encoding and processing. 
Regarding prompting strategies, approaches like Dater \cite{Dater} and DIN-SQL \cite{DIN-SQL} adopt task decomposition for better understanding with simplified queries.
Chain-of-Table \cite{Chain-of-Table} defines atomic operations for dynamic planning inspired by chain-of-thought prompting \cite{CoT}. 
Similarly, StructGPT \cite{StructGPT} leverages an iterative reading-then-reasoning approach to enhance LLM reasoning on structured data.
TableMeetsLLM \cite{TableMeetLLM} comprehensively evaluates the capabilities of LLMs in several fine-grained table understanding tasks, followed by a novel self-augmentation for effective structural prompting. 
Some other approaches explore pre-training or tuning LLMs as tabular generalists. 
TableLlama \cite{TableLlama} develops an open-source tabular LLM by fine-tuning Llama2-7B \cite{Llama2} with LongLoRA \cite{LongLoRA}.
TabPedia \cite{TabPedia} trains a visual-language model that unifies several visual understanding and reasoning tasks with a concept synergy mechanism. 
However, these strategies often focus on local step-by-step reasoning, lacking a systematic global planning perspective to effectively guide the analysis of complex tabular data.

\subsection{Table Reasoning with Symbolic Tools}

Symbolic tools have been widely utilized as assistants to produce more robust intermediate results in LLM-based table reasoning. 
Most studies adopt databases and Python to store tabular data for structured information retrieval and to execute numerical computation and other operations while interacting with LLMs. 
Studies such as Binder \cite{Binder, API-TableQA} parse the tasks into integral SQL or Python programs for further execution, incorporating LLM-assistant APIs to handle abstract code blocks. 
TabSQLify \cite{TabSQLify} generates and executes SQL queries to extract simplified sub-tables before further reasoning. 
Some approaches target boosting the code generation and execution abilities for table reasoning. 
TroVE \cite{TroVE} asks code LLMs to curate reusable high-level functions and use them to write solutions for Python execution on the table question answering and other tasks, while Self-Debugging \cite{Self-Debug} teaches LLMs to debug their predicted SQL or Python programs on text-to-SQL \cite{Spider} and other tasks. 
Furthermore, ReAcTable \cite{ReAcTable} enhances the ReAct framework by integrating a dynamic tool-use mechanism tailored for tabular data, enabling autonomous table exploration.
Recently, tool-based table reasoning research has been extended into more sophisticated tasks and scenarios. 
SheetCopilot \cite{SheetCopilot} and SpreadsheetBench \cite{SpreadsheetBench} address a novel spreadsheet manipulation task, which manipulates table analysis software like Microsoft Excel\footnote{https://www.microsoft.com/zh-cn/microsoft-365/excel} to generate step-by-step solutions for simulated execution.  
MatPlotAgent \cite{MatPlotAgent} addresses the task of scientific data visualization by designing a coding agent, which integrates \texttt{Matplotlib}\footnote{https://matplotlib.org/} package responsible for generating the code to plot figures from input tables. 
While symbolic tools enhance table reasoning, current methods typically execute code as a monolithic process, lacking the systematic, stage-oriented verification characteristic of human analysts.

\section{\textsc{PoTable}}

\begin{figure}
\begin{center}
    \includegraphics[width=250pt]{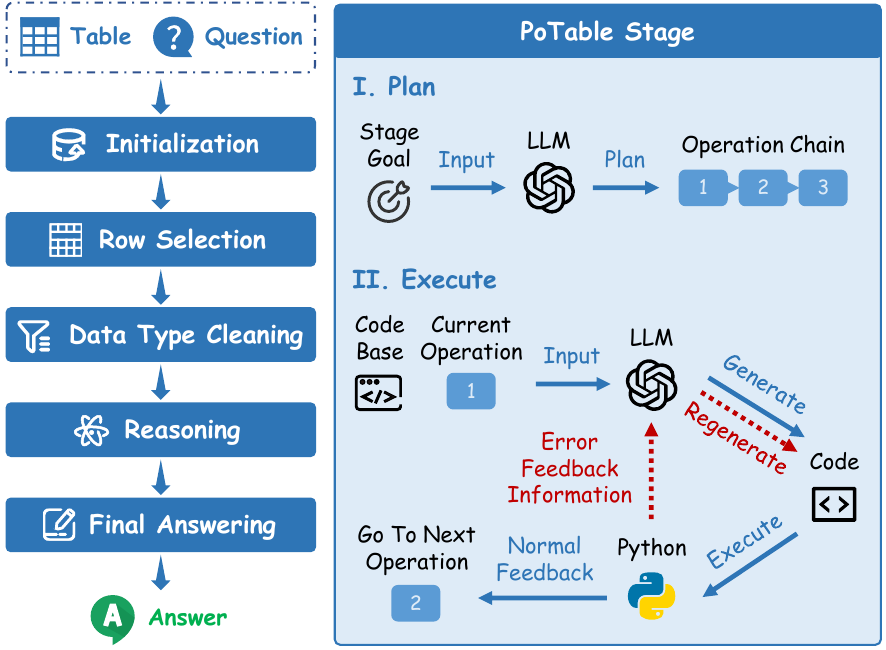}
\end{center}
\caption{Illustration of \textsc{PoTable}, a novel LLM-based table reasoning method that realizes systematic thinking. \textsc{PoTable} follows stage-oriented thinking including five analytical stages with relevant objectives and instructions: initialization, row selection, data type cleaning, reasoning and final answering. To achieve each stage-specific goal, \textsc{PoTable} integrates an LLM and a Python interpreter to conduct plan-then-execute thinking for table reasoning. }
\label{potable}
\end{figure}

In this paper, we propose \textsc{PoTable} (\textbf{P}rogramming \textbf{o}n \textbf{Table}s), a novel stage-oriented, plan-then-execute reasoning approach that integrates systematic thinking into tabular analysis, as illustrated in Fig.~\ref{potable}. 
This section first formalizes the evaluated table reasoning tasks, then provides an overview of the \textsc{PoTable} framework, and subsequently details its core components.

\subsection{Task Formulation}

This study focuses on two fundamental table reasoning tasks: \textit{table question answering (QA)} and \textit{table fact verification}. 
Formally, each sample can be represented as $(T, Q, A)$, where $T$ denotes the structured table, and $Q$ represents either a question to be answered or a statement to be verified. 
Given $T$ and $Q$, the goal is to find the answer $A$ in the table QA task. 
For table fact verification, the goal is to predict a binary label $A \in \{0, 1\}$, indicating whether the statement $Q$ is refuted or entailed by the table $T$, respectively. 

\subsection{Overview}

\textsc{PoTable} is an LLM-driven table reasoning framework that employs a stage-oriented plan-then-execute mechanism. 
Specifically, \textsc{PoTable} employs five distinct tabular analytical stages: initialization, row selection, data type cleaning, reasoning and final answering. 
Following such structured guidance, \textsc{PoTable} navigates these stages sequentially to accomplish the overall task. 
Within each stage, \textsc{PoTable} employs a collaborative plan-then-execute mechanism involving an LLM and a real-time Python interpreter. 
In the planning phase, the LLM formulates an operation chain tailored to the stage-specific objective. 
As for the executing phase, the LLM generates corresponding code for each operation, which is immediately executed by the Python interpreter. 
If the execution yields an error, the interpreter reverts to its previous state and feeds the error message back to the LLM, triggering a regeneration loop until the code executes successfully. 
Through the systematic, stage-oriented thinking, \textsc{PoTable} generates fully executable and verifiable programs, significantly reducing the risks of omitted steps, disorganized logic, and misleading conclusions. 

\subsection{Stage-oriented Thinking}

Professional human analysts inherently adopt a structured, stage-by-stage workflow when navigating complex tabular data analysis. 
Rather than relying on a single monolithic process or trial-and-error attempts, they decompose the analytical lifecycle into manageable objectives. 
Inspired by this human cognitive behavior, \textsc{PoTable} employs a stage-oriented thinking mechanism. 
This top-down structural guidance mitigates the hallucination and omission risks prevalent in direct LLM exploration by tightly constraining the operational search space within each analytical stage.

\textbf{Taxonomy of Tabular Analytical Stages. }
To seamlessly bridge LLM reasoning with precise programmatic execution, the overall table reasoning process is formalized into five distinct, sequential stages. 
In each stage, the thinking process focuses on stage-specific objectives as a part of the overall task goal, ensuring high-quality stage-wise and global table reasoning processes. 
This taxonomy is strictly derived from the fundamental requirements of programmatic tabular data analysis, covering following stages:
\begin{itemize}
    \item \textbf{Initialization}: 
    As a prerequisite, this stage bridges the textual representation of tables with the Python execution environment by parsing the raw data into a standardized \texttt{pandas.DataFrame} object. 
    \item \textbf{Row Selection}: 
    Serving as a critical data denoising step, this stage filters out redundant or irrelevant records to ensure the subsequent reasoning process focuses strictly on highly relevant records.
    For example, when counting the number of record rows under specific conditions, removing summary rows that do not present a distinct record is effective. 
    \item \textbf{Data Type Cleaning}: 
    As initialized tables typically default to noisy string formats, this stage explicitly casts target columns to appropriate programmatic types. 
    Such a cleaning process prevents execution crashes and logic failures during downstream arithmetic or conditional evaluations.
    For example, comparing two values of column ``score'' in string and integer types leads to different results.   
    \item \textbf{Reasoning}: 
    As the main process stage that will also be conducted by human analysts, this stage is indispensable for implementing flexible and useful reasoning operations towards the task question. 
    Suggested choices include sorting, arithmetic computations and other affiliated operations. 
    The completion of this stage makes it close to the final answer. 
    \item \textbf{Final Answering}: 
    While there may be multiple intermediate results, this conclusive stage logically formulates and output the final user-facing answers through a joint analysis (\textit{e.g.}, logical combination of sub-facts verification results), which is indispensable. 
\end{itemize} 
In the above stages, only the initialization stage is implemented by executing the pre-defined Python template code, while other stages will be traversed through the collaboration of the integrated LLM and the Python interpreter (integrating \texttt{pandas}\footnote{https://pandas.pydata.org/} library as a common choice of human tabular data analysts). 

\textbf{Empirical Design and Extensibility. }
The design of these five stages is primarily inspired by well-established data mining and knowledge discovery workflows \cite{KDD-1, KDD-2}, as we adapt the involved high-level concepts (\textit{e.g.}, initialization, processing, reasoning, conclusion) into concrete, executable components tailored for LLM-based programmatic table reasoning. 
Instead of claiming the five-stage selection as the only possible solution, we present it as a highly practical and empirically validated configuration. 
As demonstrated in our ablation study (Sec.~\ref{sec:abl}), this five-stage architecture provides sufficient systematic guidance to reduce misleading steps or disorganized logic without over-interference. 
Specifically, omitting the fundamental stages of row selection or data type cleaning predictably leads to semantic mismatches and execution errors, while additional new stages like column selection are not always safe, inadvertently hiding useful columns without meta information. 
Furthermore, since \textsc{PoTable} is characterized by stage-wise modular design, the analytical stages can be customized depending on the specific task complexities.

\subsection{Plan-then-Execute Reasoning}

To accomplish the entire table reasoning task by completing each stage-specific goal, \textsc{PoTable} conducts plan-then-execute reasoning through the collaboration of an integrated LLM and a Python interpreter. 
Different from conventional \textit{chain-of-thought} \cite{CoT} prompting, the plan-then-execute reasoning divides the planning phase and the executing phase to further refine the reasoning process and improve accuracy.
Specifically, \textsc{PoTable} contains a planning phase to produce a list of executable stage operations, and an executing phase to generate code and perform real-time code generation, running and feedback. 
The capabilities of LLMs in thinking decomposition and code generation are deeply promoted to improve reasoning quality. 
Additionally, this approach also enjoys the benefits of structured table memorization and precise computational results with the utilization of symbolic tools. 
The overall procedure of \textsc{PoTable} is illustrated in Algorithm~\ref{alg}.

\begin{algorithm}[t]
\renewcommand{\algorithmicrequire}{\textbf{Input:}}
\renewcommand{\algorithmicensure}{\textbf{Output:}}
\caption{\textsc{PoTable}}
\label{alg}
\begin{algorithmic}[1]
    \REQUIRE Table $T$, Question or Statement $Q$, LLM $M$, Python Interpreter $R$ and Prompts $P$
    \ENSURE Answer $A$ to the Question or Statement

    \STATE $C \gets \operatorname{initalCode}(T)$
    \STATE $R.\operatorname{executeCode}(C)$

    \FOR{$\operatorname{stage}$ $\operatorname{in}$ $\{\operatorname{``RowSel''}, \operatorname{``DtyCle''}, \operatorname{``Reason''}, \operatorname{``FinAns''}\}$}
        \STATE $\operatorname{stageBlock} \gets \operatorname{new} \operatorname{PoTableBlock}(P[\operatorname{stage}])$
        \STATE $M$.$\operatorname{clearHistory}()$
        \STATE $\operatorname{operationList} \gets \operatorname{stageBlock.plan}(T, Q, M)$
        \FOR{$o$ $\operatorname{in}$ $\operatorname{operationList}$}
            \STATE $M$.$\operatorname{clearHistory}()$
            \STATE $c\gets \operatorname{stageBlock.codeGen}(T, Q, M, C, o)$
            \STATE $\operatorname{errorCnt} \gets 0$
            \WHILE {$\operatorname{catchError}(R.\operatorname{executeCode(code)})$ $\operatorname{as}$ $e$ $\operatorname{and}$ $\operatorname{errorCnt}<\operatorname{upLimit}$}
                \STATE $R.\operatorname{restart().executeCode(C)}$
                \STATE $c\gets\operatorname{stageBlock.codeRegen}(T, Q, M, C, o, c, e)$
                \STATE $\operatorname{errorCnt}\gets\operatorname{errorCnt}+1$
            \ENDWHILE
            \STATE $C \gets C + \operatorname{code}$
            \STATE $T \gets R.\operatorname{getCurrentStatus}(T)$
        \ENDFOR
    \ENDFOR
    \STATE $A\gets R.\operatorname{getProgramOutput}()$
    \RETURN $A$
\end{algorithmic}
\end{algorithm}

\begin{figure}
\begin{center}
    \includegraphics[width=250pt]{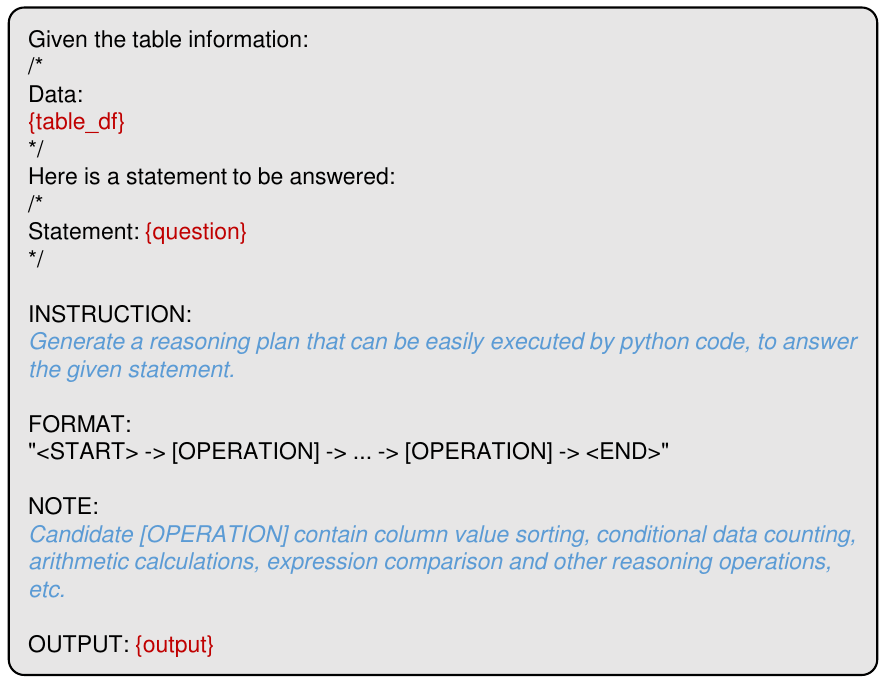}
\end{center}
\caption{An example of a planning prompting template in the reasoning stage for the WikiTQ dataset, whose texts of instruction and note will be changed in different stages. }
\label{planning}
\end{figure}

\subsubsection{Planning Phase}

Inspired by \textit{chain-of-thought} \cite{CoT} prompting, \textsc{PoTable} formulates the operation chain guided by the corresponding stage-specific goal and instruction through the integrated LLM.
An example of a stage planning prompting template is shown in Fig.~\ref{planning}. 
Specifically, the prompting template contains the tabular data in Markdown format and the task question, followed by an instruction part integrating the stage-specific goal, output format and suggestion notes. 
This guides the LLM to generate the operation chain by concentrating on the stage-specific goal rather than solely on the overall task objective.
To effectively prompt the LLM, we adopt a few-shot learning strategy \cite{GPT3} by designing three custom examples for the planning phase.
Consequently, \textsc{PoTable} maintains a balance between structure and autonomy within systematic thinking, since the candidate operations remain open-world. 
More details concerning instruction prompts of the planning phase for all stages can be accessed in our code repository.  

\subsubsection{Executing Phase}

\begin{figure}
\begin{center}
    \includegraphics[width=250pt]{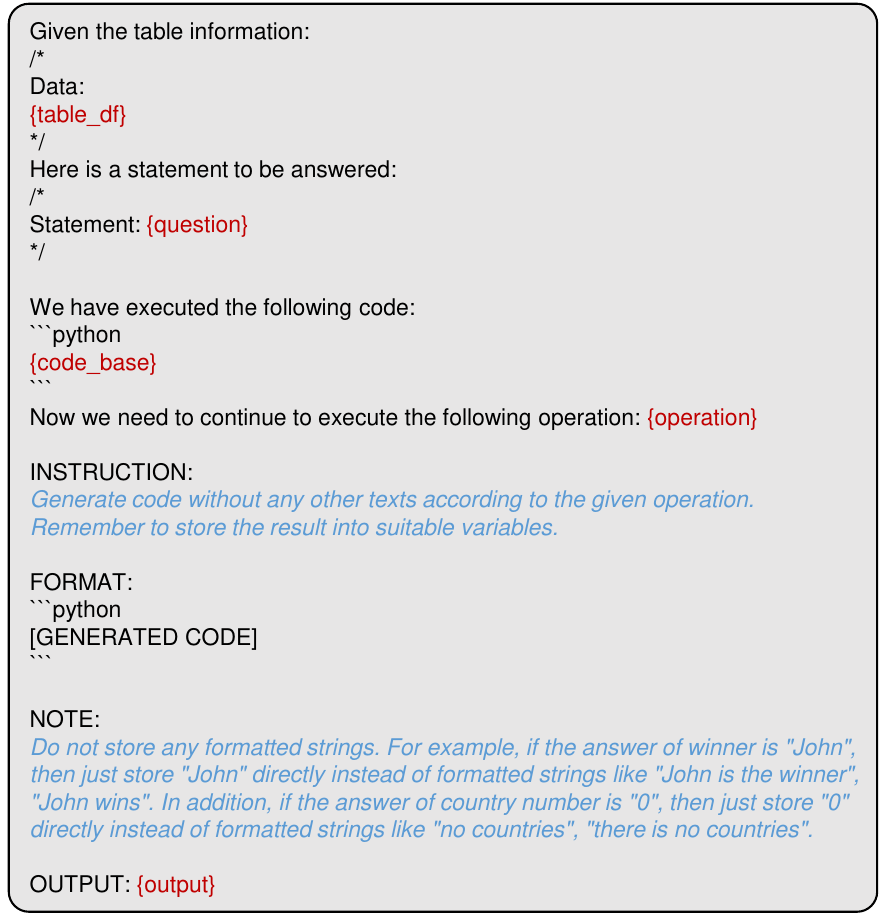}
\end{center}
\caption{An example of an executing prompting template in the reasoning stage for the WikiTQ dataset, whose texts of instruction and note will be changed in different stages. }
\label{executing}
\end{figure}

After stage-specific planning, the generated operations will be executed sequentially to realize the stage-oriented thinking. 
Rather than relying on direct LLM-based queries, which are prone to numerical reasoning errors, the executing phase operates through real-time Python code generation, running and feedback.
Specifically, the following three processes are executed for each operation:
\begin{itemize}
    \item \textbf{Code Generation}: The LLM generates suitable code for each operation based on the existing code base. 
    Formally, given the operation $o$ and the previously generated code base $C$, the LLM is required to generate Python code for $o$, which must adapt to the continuous execution state of code base $C$, denoted as $c=\operatorname{LLM}_{\operatorname{Gen}}(T, Q, C, o)$.
    \item \textbf{Real-time Running}: The generated code $c$ is executed immediately through the Python interpreter $R$ based on the status following the last execution. During this process, the execution may return feedback information, and the interpreter's pre-execution and post-execution states are recorded. 
    \item \textbf{Feedback Processing}: If the real-time execution returns erroneous information $e$, (\textit{e.g.}, runtime error), the LLM will regenerate the code using the feedback, denoted as $c_{\operatorname{new}}=\operatorname{LLM}_{\operatorname{Regen}}(T,Q,C,o,c,e)$. 
    In addition, the Python interpreter $R$ will be rolled back to the state before executing $c$, initiating a new execution loop.
\end{itemize}
An example of an operation executing prompting template is shown in Fig.~\ref{executing}.
In the final answering stage, we adopt few-shot prompting with three self-made examples to obtain the code to print out the answer, formatting the output strictly to align with dataset-specific evaluation metrics.
In contrast, we adopt zero-shot prompting to generate the code in other stages, at which modern LLMs are already highly proficient in standard programmatic manipulations.

Ultimately, the final answer is derived from the full program execution of the sequential stage codes rather than direct LLM-generated text. 
Throughout the entire process, \textsc{PoTable} demonstrates a high degree of alignment with the cognitive behavior of a professional data analyst, offering key advantages concerning accuracy and explainability.

% \subsection{Summarization}

% Throughout the entire process, \textsc{PoTable} demonstrates a high degree of alignment with the cognitive behavior of a distinguished analyst, offering two key advantages: 
% \begin{itemize}
%     \item \textbf{Accuracy}: \textsc{PoTable} plans coherent operation chains under precise stage-specific goals. In the shorter stage operation chain than the one of the entire task, there is less possibility for \textsc{PoTable} to generate omitted steps or misleading results. 
%     The stage-oriented thinking helps enhance the reliability of final answers. 
%     \item \textbf{Explainability}: Following stage-oriented plan-then-execute reasoning, \textsc{PoTable} generates executable programs with precise stage boundaries along with the reasoning results. 
%     Accordingly, it is easy to review the completeness and accuracy of the procedure.
% \end{itemize}
% Consequently, \textsc{PoTable} produces outstanding reasoning results with highly accurate, steply commented and completely executable Python programs. 
% As they correspond to clear stage-specific operations and have experienced real-time execution and validation, the effectiveness of systematic thinking has been fully demonstrated.

\section{Experiments}

\subsection{Experimental Setup}

\textbf{Datasets}. We conduct experiments on datasets from two public table reasoning benchmarks: \textbf{WikiTQ} \cite{WikiTQ} and \textbf{TabFact} \cite{TabFact}. 
WikiTQ is a table QA benchmark that requires answering the question using short text spans derived from a given Wikipedia table. 
We evaluate all approaches on the official development (dev) set under split-1 and the unseen test set.
These two evaluation sets are completely disjoint, ensuring a rigorous evaluation of the compared approaches on table QA.
TabFact is a table fact verification benchmark that requires judging whether a given statement is entailed by a provided Wikipedia table. 
We use the small test set following most prior studies, and include the complex test set that involves a more challenging verification process, to thoroughly assess the reasoning capabilities of all compared approaches.
Additionally, we categorize the difficulty of the evaluated questions or statements as either \textit{simple} or \textit{complex}.
For WikiTQ, a question with a length less than 50 is labeled as \textit{simple}, while longer ones are considered \textit{complex}. 
For TabFact, we adopt the official difficulty labels for all statements.
We also classify tables by size into \textit{small} (1–49 content cells), \textit{medium} (50–99 content cells), and \textit{large} (100 or more content cells).
Detailed statistics of the evaluation datasets are summarized in Table~\ref{statistics}.
Regarding evaluation metrics, we report the official denotation accuracy for WikiTQ and binary classification accuracy for TabFact.

\begin{table}[]
\caption{Statistics of the evaluation datasets: WikiTQ (\textbf{D}, \textbf{T} denotes development and test set respectively) and TabFact (\textbf{S}, \textbf{C} denotes small and complex test set respectively), with group information of difficulty (\textbf{S}, \textbf{C} denotes simple and complex respectively) and table size (\textbf{S}, \textbf{M}, \textbf{L} denotes small, medium and large respectively).}
\label{statistics}
\begin{center}
\begin{tabular}{cccc}
\hline
\multirow{2}{*}{\textbf{Dataset}} & \multirow{2}{*}{\textbf{\# All}} & \textbf{Difficulty} & \textbf{Table Size} \\
 &  & \# S / \# C & \# S / \# M / \# L \\ \hline
\textbf{WikiTQ (D)} & 2,831 & 1,297 / 1,534 & 573 / 976 / 1,282 \\
\textbf{WikiTQ (T)} & 4,344 & 1,993 / 2,351 & 651 / 1,737 / 1,956 \\
\textbf{TabFact (S)} & 2,024 & 1,005 / 1,019 & 574 / 839 / 611 \\
\textbf{TabFact (C)} & 8,308 & 0 / 8,308 & 3,037 / 3,151 / 2,420 \\ \hline
\end{tabular}
\end{center}
\end{table}

\textbf{Backbones}. We choose two representative LLMs as the backbones of \textsc{PoTable} and other baseline approaches in our experiments. 
Specifically, we choose GPT-4o-mini (2024-07-18)\footnote{https://openai.com/index/gpt-4o-mini-advancing-cost-efficient-intelligence/} (\textbf{GPT}) as the closed-source LLM, which is competent and cost-efficient to cover a wide range of downstream tasks. 
In addition, we choose Llama-3.1-70B-Instruct\footnote{https://ai.meta.com/blog/meta-llama-3-1/} (\textbf{LLAMA}) as the open-source LLM, which presents strong reasoning abilities among released LLMs. 

\textbf{Baselines}. To reveal the effectiveness of \textsc{PoTable}, we select four competitive LLM-based approaches as the evaluation baselines over the two table reasoning tasks. 
\textbf{Binder} \cite{Binder} is a neural-symbolic framework that parses the task into a specific program and then executes the program, binding LLM as a unified API to extend its grammar coverage to tackle the inexecutable code. 
\textbf{Dater} \cite{Dater} first decomposes the table into sub-evidence by LLM-based task-relevant column and row selection, and then decomposes the question into simpler sub-questions through intermediate SQL generation, followed by a joint reasoning stage with simplified tables and questions. 
\textbf{Chain-of-Table} \cite{Chain-of-Table} prepares several commonly used atomic operations and allows the LLM to dynamically plan the next operation from the candidate pool, forming an operation chain to process the table with pre-defined code to simplify the table for the final LLM querying.
\textbf{TabSQLify} \cite{TabSQLify} leverages text-to-SQL technologies to decompose the table into sub-tables, and then conduct chain-of-thought reasoning to generate the answers. 
All these baselines are LLM prompt-based methods instead of conventional fine-tuning for a fair comparison, which are implemented based on the released official code. 

\textbf{Implementation Details}. 
In LLM parameter setting of \textsc{PoTable}, we set the value of temperature to 0.1, top\_p to 0.9, max\_tokens to 2,048 and n\_samples to 1. 
For the baselines, we only fix n\_samples to 1 to ensure a fair comparison of generation performance. 
For other parameters, we retain their originally proposed settings, as the targeted operations and deployed modules vary across different approaches.
In addition, we carefully design prompting templates for the planning and executing phases of each stage.
We prepare three few-shot prompting examples for WikiTQ and TabFact, respectively, covering both operation planning and final answer code generation. 
To access more details, please refer to our code repository at: \href{https://github.com/Double680/PoTable}{https://github.com/Double680/PoTable}.

\begin{table*}[h]
\caption{Comparative accuracy results (\%) of LLM-based table reasoning approaches over WikiTQ (\textbf{D}), WikiTQ (\textbf{T}), TabFact (\textbf{S}) and TabFact (\textbf{C}) on GPT-4o-mini (GPT) and Llama-3.1-70B-Instruct (LLAMA). In each group, the best result is marked in \textbf{bold} and the second-best result is \underline{underlined}, while the absolute accuracy improvement of \textsc{PoTable} over the runner-up is recorded below the accuracy value. }
\label{main}
    \begin{center}
    \begin{tabular}{ccccccccccc}
    \hline
    \multirow{2}{*}{\textbf{Approach}} & \multicolumn{2}{c}{\textbf{WikiTQ (D)}} & & \multicolumn{2}{c}{\textbf{WikiTQ (T)}} & & \multicolumn{2}{c}{\textbf{TabFact (S)}} & & \textbf{TabFact (C)} \\ \cline{2-3} \cline{5-6} \cline{8-9} \cline{11-11}
     & GPT & LLAMA & & GPT & LLAMA & & GPT & LLAMA & & GPT \\ \hline
    Binder \cite{Binder} & \underline{59.20} & 50.65 & & \underline{58.86} & 50.51 & & \underline{84.63} & 78.16 & & 76.73 \\
    Dater \cite{Dater} & 56.76 & 42.78 & & 58.33 & 43.53 & & 83.99 & 81.57 & & \underline{79.02} \\
    Chain-of-Table \cite{Chain-of-Table} & 56.64 & \underline{62.39} & & 55.60 & \underline{62.22} &  & 84.24 & \underline{85.62} & & 75.10 \\
    TabSQLify \cite{TabSQLify} & 56.87 & 55.51 & & 57.02 & 55.78 & & 78.75 & 70.70 & & 71.99 \\ \hline
    \textsc{PoTable} & \begin{tabular}[c]{@{}c@{}}\textbf{63.58}\\ (+4.38) \end{tabular} & \begin{tabular}[c]{@{}c@{}}\textbf{65.10}\\ (+2.71)\end{tabular} &  & \begin{tabular}[c]{@{}c@{}}\textbf{64.73}\\ (+5.87)\end{tabular} & \begin{tabular}[c]{@{}c@{}}\textbf{65.56}\\ (+3.34)\end{tabular} & & \begin{tabular}[c]{@{}c@{}}\textbf{88.93}\\ (+4.30)\end{tabular} & \begin{tabular}[c]{@{}c@{}}\textbf{87.06}\\ (+1.44)\end{tabular} & & \begin{tabular}[c]{@{}c@{}}\textbf{82.70}\\ (+3.68)\end{tabular} \\ \hline
    \end{tabular}
    \end{center}
\end{table*}

\subsection{Quantitative Results}

We conduct experiments to compare \textsc{PoTable} with other baselines over WikiTQ (D), WikiTQ (T), TabFact (S) and TabFact (C) using GPT and LLAMA backbones (excluding LLAMA on TabFact (C) due to high token costs). 
Quantitative results are presented in Table~\ref{main}, revealing that our \textsc{PoTable} significantly outperforms all other baselines across all evaluation datasets.
In particular, GPT-based \textsc{PoTable} achieves an absolute accuracy improvement of over 4.3\% compared to the runner-ups on the first three evaluation datasets. 
Furthermore, on the larger and more complex TabFact (C), it yields a 3.68\% accuracy improvement over the second-best approach.

Specifically, we analyze the comparative results in terms of evaluation datasets, baselines and backbones.
According to the results, Binder achieves the second-best performance among GPT-based baselines on all datasets except TabFact (C).
As Binder relies heavily on full program generation, we can infer that GPT-based Binder benefits significantly from the LLM's capability in code generation.
It handles reasoning tasks of standard difficulty well, yet suffers a performance drop when tackling more complex tasks.
Furthermore, replacing the backbone with an LLM of lower coding ability negatively impacts the results.
For LLAMA-based approaches, Chain-of-Table consistently ranks as the runner-up despite its constrained operation pool for dynamic selection. 
Notably, when switching to the GPT backbone, its accuracy drops by approximately 6\% on WikiTQ and 1.4\% on TabFact (S).
We can infer that LLAMA possesses a stronger planning and reasoning capability, as Chain-of-Table heavily depends on the LLM's ability to plan and decompose operations rather than generate code, since the code for all operations is predefined.
Moreover, for Dater and TabSQLify, the performance gap between the GPT and LLAMA backbones fluctuates significantly across different evaluation datasets.
This variability suggests that their mechanisms for sub-table extraction and decomposed reasoning lack robustness across different LLMs.

In comparison, \textsc{PoTable} better promotes the capabilities of LLMs in code generation and operation planning with mitigated inter-stage influence. 
Since the entire reasoning process adheres to well-defined analytical stages, \textsc{PoTable} can concentrate on stage-specific goals to conduct planning and code generation, with less possibility of omitted steps or misleading results.
In addition, the plan-then-execute reasoning method mitigates conflicts arising from the blurred boundaries between planning and executing. 
As a result, the stage-oriented plan-then-execute reasoning leads \textsc{PoTable} to the best accuracy in both simple and complex scenarios, indicating its effectiveness for table reasoning.

\subsection{Group Analysis}

To conduct a deeper analysis of \textsc{PoTable}, we re-evaluate its accuracy across different categories of task difficulty and table size to provide a fine-grained assessment.
Detailed grouped results are reported in Table \ref{analysis}. 
As observed, the accuracy on simple task groups is consistently higher than that on complex task groups, confirming that increased task complexity predictably leads to a performance decline.
It suggests that integrating explicit parallel sub-task decomposition into \textsc{PoTable} could serve as a potential avenue for future improvement.
Furthermore, the reasoning performance on small and medium tables is slightly better than that on large tables.
Overall, \textsc{PoTable} exhibits relative robustness to varying table sizes in standard tabular data analysis.

\begin{table}[]
\caption{Accuracy results (\%) of \textsc{PoTable} on different groups in task difficulty as simple (S) and complex (C) and different groups in table size as small (S), medium (M) and large (L). }
\label{analysis}
\begin{center}
\scalebox{0.96}{
\begin{tabular}{cccccccc}
\hline
\multirow{2}{*}{\textbf{LLM}} & \multirow{2}{*}{\textbf{Dataset}} & \multicolumn{2}{c}{\textbf{Difficulty}} & \textbf{} & \multicolumn{3}{c}{\textbf{Table Size}} \\ \cline{3-4} \cline{6-8}
 &  & S & C &  & S & M & L \\ \hline
\multirow{4}{*}{GPT} & WikiTQ (D) & 67.77 & 60.04 & \textbf{} & 63.00 & 65.57 & 62.32 \\
 & WikiTQ (T) & 68.99 & 61.12 & \textbf{} & 70.20 & 66.21 & 61.61 \\
 & TabFact (S) & 90.65 & 87.24 &  & 90.59 & 88.44 & 88.05 \\ 
 & TabFact (C) & - & 82.70 &  & 82.32 & 84.04 & 81.45 \\ \hline
\multirow{3}{*}{LLAMA} & WikiTQ (D) & 68.00 & 62.65 &  & 68.41 & 67.52 & 61.78 \\
 & WikiTQ (T) & 68.89 & 62.74 &  & 71.27 & 67.99 & 61.50 \\
 & TabFact (S) & 88.96 & 85.18 &  & 85.71 & 87.84 & 87.23 \\ \hline
\end{tabular}
}
\end{center}
\end{table}

\subsection{Ablation Study}
\label{sec:abl}
The overall table reasoning process of \textsc{PoTable} comprises five distinct stages.
To validate the rationality of these selected stages, we conduct an ablation study by adopting different stage divisions, comparing the original GPT-based \textsc{PoTable} with the following four settings:
\begin{itemize}
    \item \textbf{Only Reasoning} (\textit{only} Reason): This setting removes explicit stage divisions, forcing \textsc{PoTable} to rely solely on a single overarching reasoning stage. 
    \item \textbf{Removing Row Selection} (\textit{w/o} Row Sel.): This setting skips the removal of redundant rows prior to downstream processing. In previous studies, this is often considered a crucial sub-table extraction step.
    \item \textbf{Removing Data Type Cleaning} (\textit{w/o} Dty. Cle.): This setting bypasses column type transformations. Since initialized tables default to string formats, omitting this operation increases the risk of execution errors.
    \item \textbf{Adding Column Selection} (\textit{w/} Col. Sel.): Rather than removing a stage, we introduce an explicit column selection phase prior to processing—a common sub-table extraction operation in existing literature.
\end{itemize}

\begin{figure*}
\begin{center}
    \includegraphics[width=520pt]{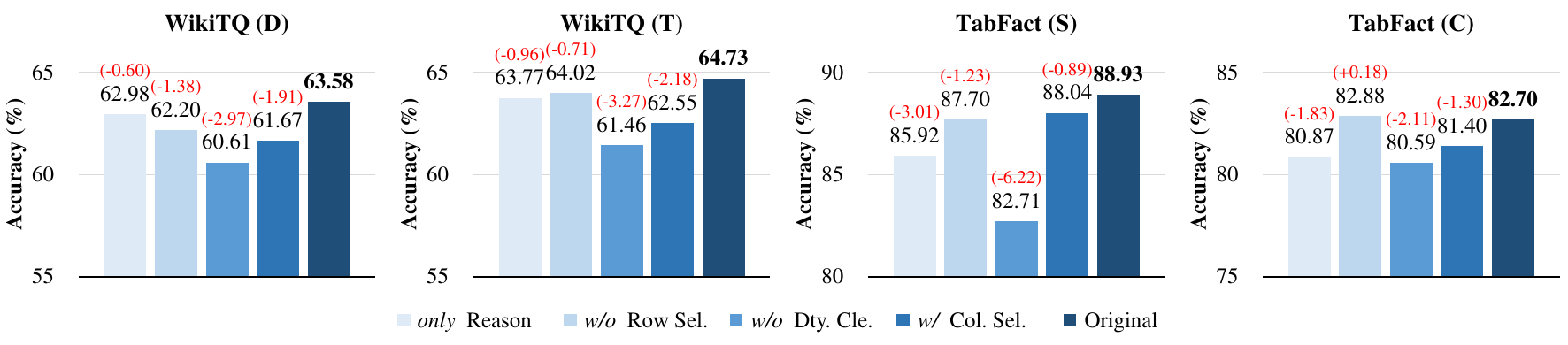}
\end{center}
\caption{Accuracy results (\%) in the ablation study of the different stage division settings employed in \textsc{PoTable} with GPT-4o-mini on four evaluation datasets of WikiTQ and TabFact. These settings contain only reasoning (\textit{only} Reason), removing row selection (\textit{w/o} Row Sel.), removing data type cleaning (\textit{w/o} Dty. Cle.), adding column selection (\textit{w/} Col. Sel.) and the original setting (Original). 
The best results are marked in \textbf{bold}, while the accuracy differences in all settings are recorded in red. }
\label{ablation}
\end{figure*}

% \begin{table}[]
% \caption{Fine-grained accuracy results (\%) in the ablation study grouped by different task difficulty as simple (S) and complex (C) in GPT-based \textsc{PoTable} on WikiTQ (T) and TabFact (S). }
% \label{ablation-diff}
% \begin{center}
% \begin{tabular}{cccccc} \hline
% \multirow{2}{*}{\textbf{Setting}} & \multicolumn{2}{c}{\textbf{WikiTQ (T)}} &  & \multicolumn{2}{c}{\textbf{TabFact (S)}} \\ \cline{2-3} \cline{5-6}
%  & S & C &  & S & C \\ \hline
% only Reason & 66.98 & 60.19 &  & 85.87 & 85.97 \\
% w/o Row Sel. & 67.74 & 60.87 &  & 88.66 & 86.75 \\
% w/o Dty. Cle. & 65.48 & 58.06 &  & 80.70 & 84.69 \\
% w/ Col. Sel. & 66.18 & 59.46 &  & 88.96 & 87.14 \\
% \textbf{Original} & \textbf{68.99} & \textbf{61.12} &  & \textbf{90.65} & \textbf{87.24} \\ \hline
% \end{tabular}
% \end{center}
% \end{table}

The overall results of the ablation study on different stage settings are shown in Fig.~\ref{ablation}. 
We observe that the original GPT-based \textsc{PoTable} outperforms all ablated settings in almost all evaluation datasets. 
Specifically, compared to the original configuration, the \textit{only Reason} setting scores approximately 0.6\%–1\% lower on WikiTQ, 3.01\% lower on TabFact (S), and 1.83\% lower on TabFact (C).
While it remains competitive on WikiTQ, it suffers a significant accuracy drop on TabFact.
From the task perspective, WikiTQ involves retrieving explicit key information to answer direct questions, whereas TabFact requires verifying statements that often entail complex, multi-hop reasoning under implicit conditions.
Consequently, structured stage-oriented thinking plays a pivotal role in \textsc{PoTable}, particularly in complex scenarios like TabFact. 
In addition, the results of ``\textit{w/o} Row Sel.'' and ``\textit{w/o} Dty Cle.'' demonstrate their importance as expected. 
Notably, \textit{w/o} Dty Cle.'' consistently yields the lowest accuracy, indicating that without explicit stage division, the framework may fail to handle crucial format conversions.  
However, we observe a minor exception on TabFact (C), where the ``\textit{w/o} Row Sel.'' setting marginally outperforms the original framework. 
As no explicit metadata is provided, LLMs may fall short in global data awareness when facing complex scenarios, leading to aggressive row filtering that inadvertently removes nuanced rows. 
Such a conclusion aligns with the degraded performance of the ``\textit{w/} Col. Sel.'' setting, despite its widespread adoption in previous studies. 
Without tabular metadata, eliminating seemingly irrelevant columns introduces a high risk of accidentally discarding vital contextual data in LLM-based table reasoning.

\begin{table}[]
\caption{Fine-grained accuracy results (\%) in the ablation study grouped by different task difficulty as simple (S) and complex (C), and different table sizes as small (S), medium (M) and large (L) in GPT-based \textsc{PoTable} on WikiTQ (T) and TabFact (S). }
\label{ablation-fine}
\begin{center}
\begin{tabular}{cccccccc}
\hline
\multirow{2}{*}{\textbf{Dataset}} & \multirow{2}{*}{\textbf{Setting}} & \multicolumn{2}{c}{\textbf{Difficulty}} & \textbf{} & \multicolumn{3}{c}{\textbf{Table Size}} \\ \cline{3-4} \cline{6-8}
 &  & S & C &  & S & M & L \\ \hline
\multirow{5}{*}{\begin{tabular}[c]{@{}c@{}}WikiTQ\\ (T)\end{tabular}} & \textit{only} Reason & 66.98 & 60.19 &  & 68.82 & 64.59 & 61.35 \\
 & \textit{w/o} Row Sel. & 67.74 & 60.87 &  & 69.74 & 65.28 & 60.99 \\
 & \textit{w/o} Dty. Cle. & 65.48 & 58.06 &  & 66.67 & 63.50 & 57.92 \\
 & \textit{w/} Col. Sel. & 66.18 & 59.46 & \textbf{} & 67.28 & 65.05 & 58.74 \\
 & \textbf{Original} & \textbf{68.99} & \textbf{61.12} &  & \textbf{70.20} & \textbf{66.21} & \textbf{61.61} \\ \hline
\multirow{5}{*}{\begin{tabular}[c]{@{}c@{}}TabFact\\ (S)\end{tabular}} & \textit{only} Reason & 85.87 & 85.97 &  & 85.89 & 86.53 & 85.11 \\
 & \textit{w/o} Row Sel. & 88.66 & 86.75 &  & 87.80 & 87.37 & \textbf{88.05} \\
 & \textit{w/o} Dty. Cle. & 80.70 & 84.69 &  & 83.80 & 83.55 & 80.52 \\
 & \textit{w/} Col. Sel. & 88.96 & 87.14 &  & 89.02 & 87.49 & 87.89 \\
 & \textbf{Original} & \textbf{90.65} & \textbf{87.24} &  & \textbf{90.59} & \textbf{88.44} & \textbf{88.05} \\ \hline
\end{tabular}
\end{center}
\end{table}

Furthermore, we also report the fine-grained group results of the ablation study in GPT-based \textsc{PoTable} on WikiTQ (T) and TabFact (S), as illustrated in Table~\ref{ablation-fine}.  
The original \textsc{PoTable} reaches the best results in all different difficulty groups and table size groups, strengthening the conclusions from the above ablation results and analysis. 

\begin{table*}[]
\caption{Efficiency results on TabFact (S) for GPT-based methods. 
For the three baselines, we compared the results of single LLM generation (Single) and default LLM generation (Default) following their claimed settings in the article. }
\label{efficiency}
\begin{center}
\begin{tabular}{ccccl}
\hline
\multirow{2}{*}{\textbf{Approach}} & \multicolumn{2}{c}{\textbf{Accuracy}} & \multirow{2}{*}{\textbf{\begin{tabular}[c]{@{}c@{}}\# Generation\\ (Default)\end{tabular}}} & \multirow{2}{*}{\textbf{Details}} \\ \cline{2-3}
 & Single & Default &  &  \\ \hline
Binder \cite{Binder} & 84.63 & 85.13 & 50 & SQL Generation: 50 \\ \hline
Dater \cite{Dater} & 83.99 & 86.81 & 100 & \begin{tabular}[c]{@{}l@{}}Decomposition Generation: 40, Cloze Generation: 20,\\ SQL Generation: 20, Query: 20\end{tabular} \\ \hline
Chain-of-Table \cite{Chain-of-Table} & 84.24 & 85.23 & $\leq$22 & \begin{tabular}[c]{@{}l@{}}Dynamic Planning: $\leq$4 (3.74 on average), \\ Args Generation: $\leq$17 (16.09 on average), Query: 1\end{tabular} \\ \hline
\begin{tabular}[c]{@{}c@{}}\textsc{PoTable}\\ (\textit{only} Reason)\end{tabular} & \multicolumn{2}{c}{85.92} & $\leq$6 & \begin{tabular}[c]{@{}l@{}}Planning: 1, Code Generation: $\leq 4$ (3.72 on average),\\ Re-Generation: $\leq$1 (less than 1 on average)\end{tabular} \\ \hline
\textsc{PoTable} & \multicolumn{2}{c}{\textbf{88.93}} & $\leq$10 & \begin{tabular}[c]{@{}l@{}}Planning: 3, Code Generation: $\leq 6$ (5.60 on average),\\ Re-Generation: $\leq$1 (less than 1 on average) \end{tabular} \\ \hline
\end{tabular}
\end{center}
\end{table*}

\subsection{Efficiency Analysis}

To further show that the improvements in \textsc{PoTable} do not come from sacrificing computational overhead, we analyze the efficiency of GPT-based \textsc{PoTable} and three representative approaches by evaluating the number of required LLM-based generations in TabFact (S). 
Analytical results are presented in Table \ref{efficiency}. 
The baseline approaches require multiple LLM generations for their intermediate steps (marked as Default), whereas \textsc{PoTable} adopts a single-generation strategy by default. 
In Binder and Dater, the generation counts are fixed while the ones of Chain-of-Table and \textsc{PoTable} fluctuate dynamically, so we report the empirical average counts of each module for estimation. 
We observe that while multiple generations yield some improvements for the baselines, they inherently demand significantly higher generation counts, resulting in substantial token consumption. 
In contrast, \textsc{PoTable} strictly adheres to a single-generation approach, requiring far fewer inference steps while still achieving the best performance among all baselines. 
In addition, the difference in generation counts between the original \textsc{PoTable} and the ``\textit{only} Reason'' setting is quite small, whereas the accuracy gap exceeds 3\%. 
This confirms that the significant performance gains stem from the stage-oriented plan-then-execute reasoning paradigm rather than brute-force multiple generations, thoroughly validating the efficiency of our \textsc{PoTable}.

\begin{figure*}
\begin{center}
    \includegraphics[width=520pt]{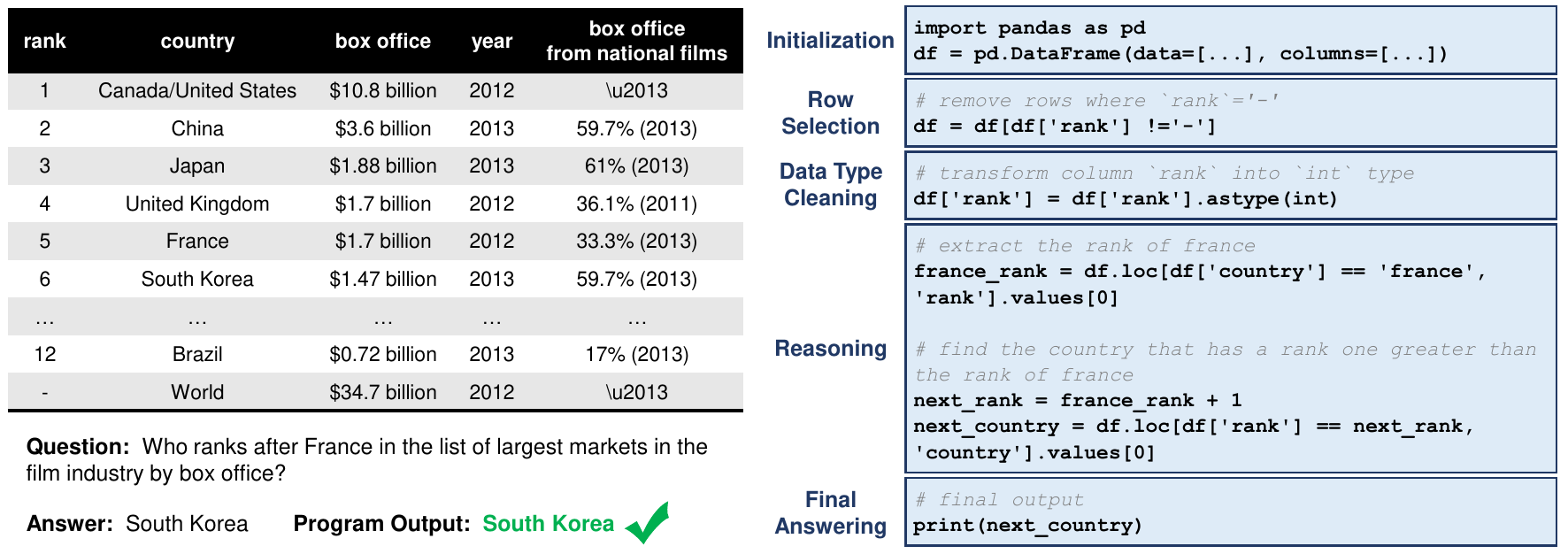}
\end{center}
\caption{A case study of an evaluated sample from WikiTQ (T) with its generated Python program and output answer. The program is fully executable with precise stage boundaries, making it easier to review and analyze the reasoning process.}
\label{case}
\end{figure*}

\subsection{Case Study}
\begin{figure*}
\begin{center}
    \includegraphics[width=520pt]{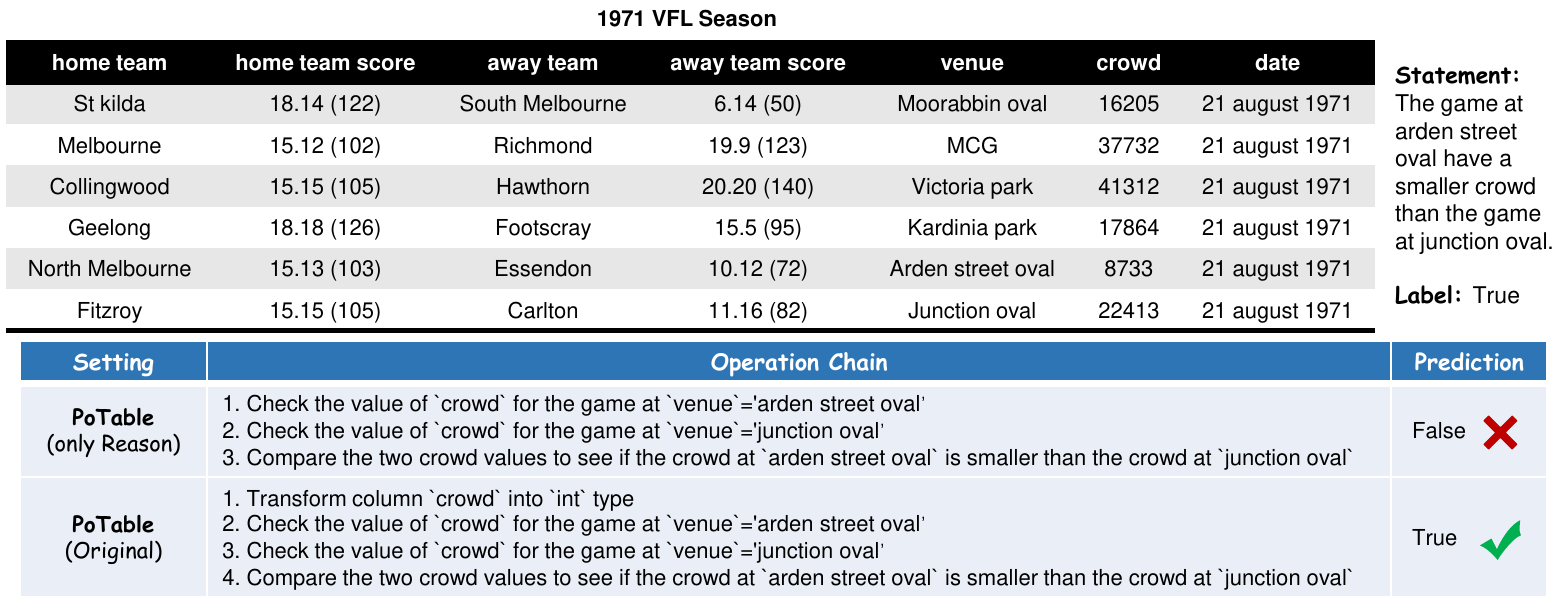}
\end{center}
\caption{A case study of an evaluated sample from TabFact (C) with the generated operation chains of \textsc{PoTable} (\textit{only} Reason) and \textsc{PoTable}. Through stage-oriented thinking, \textsc{PoTable} mitigates missing or misleading steps and achieves better results. }
\label{comparison}
\end{figure*}

We conduct case analysis to further showcase the effectiveness of \textsc{PoTable}.
We first present a case study from the WikiTQ (T) dataset, illustrating the generated Python program and final output of \textsc{PoTable}, as shown in Fig.~\ref{case}.
To answer the question, \textsc{PoTable} employs stage reasoning, systematically deriving the answer through a fully complete program.
The program is step-wise commented and entirely executable, with precise boundaries between analytical stages. 
This transparency allows us to easily review the reasoning trajectory and accurately diagnose the root causes of specific outputs.
Furthermore, Fig.~\ref{comparison} contrasts the original \textsc{PoTable} with the ablated \textit{only Reason} setting using an evaluated sample from TabFact (C).
Since the \textit{only Reason} setting lacks explicit stage divisions, it is forced to plan the operation chain based solely on the overarching task objective.
As a result, it frequently overlooks crucial intermediate steps involving data type unification, negatively impacting the final result.
More importantly, the cause of the erroneous result remains difficult to reveal from the operation chain.  
In contrast, the original \textsc{PoTable} formulates plans under focused, stage-specific targets, thereby mitigating the risk of omitted steps and misleading conclusions. 
Overall, these successful cases and analysis strongly highlight the superior accuracy and explainability of \textsc{PoTable}. 

\subsection{Failure Analysis}

\begin{figure*}
\begin{center}
    \includegraphics[width=520pt]{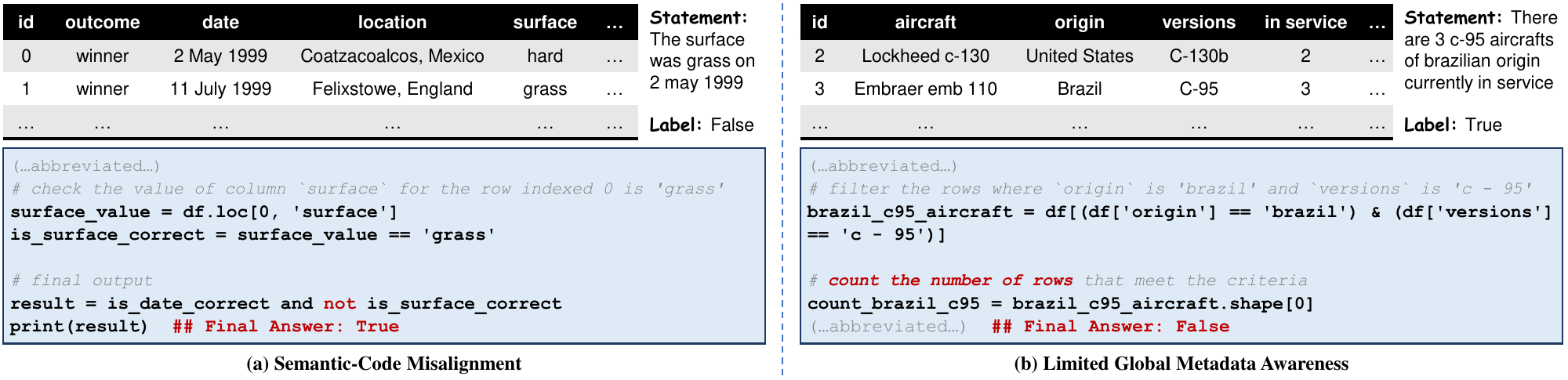}
\end{center}
\caption{Error case analysis of \textsc{PoTable} from TabFact (S), concerning (a) semantic-code misalignment and (b) limited global metadata awareness. }
\label{failure}
\end{figure*}

Despite its promising results on two public benchmarks, \textsc{PoTable} occasionally yields suboptimal results. 
To reveal the potential limitations, we re-evaluate the results on TabFact (S) and identify some common failure patterns.
Qualitative analysis of failure cases in Fig.~\ref{failure} reveals representative errors caused by the inherent backbone limitations rather than a breakdown of our stage-oriented framework. 
Fig.~\ref{failure}(a) illustrates a semantic-code misalignment: the LLM redundantly inserts a ``not'' operator during the final answering stage, creating a double negative and flipping the final result to ``True''. 
This suggests the confusion of natural language semantics with strict code logic. 
Furthermore, Fig.~\ref{failure}(b) highlights the limited global metadata awareness from the LLM's perspective: it incorrectly counts relevant rows instead of directly querying the target ``in service''. 
From a common-sense perspective, both approaches appear logically plausible, yet operating without explicit metadata makes the LLM blindly select the inappropriate execution path.
Crucially, the stage-oriented \textsc{PoTable} operates exactly as intended in both scenarios. 
These failures underscore the inherent challenges LLMs face in strict semantic-code translation and holistic table comprehension, which could be mitigated and improved by integrating strict logic verification or global schema-checking modules.

% \textbf{Semantic-Code Misalignment}. As shown in Fig.~\ref{fig:failure_cases}(a), the reasoning plan is flawless, but the LLM confuses natural language semantics with code logic during the final answering stage. Although the intermediate boolean variable correctly evaluates to \texttt{False} (since the surface is not grass), the LLM redundantly inserts a \texttt{not} operator in the final return statement. This semantic hallucination inadvertently creates a double negative, flipping the final result to \texttt{True}.

% \textbf{Lack of Global Data Awareness}. Fig.~\ref{fig:failure_cases}(b) highlights an error caused by insufficient comprehension of the table's schema. To verify the aircraft quantity, the LLM incorrectly counts the relevant rows instead of directly querying the provided ``in service'' column. From a common-sense perspective, both approaches appear logically plausible. However, operating without explicit metadata or global data awareness, the LLM blindly selects the inappropriate execution path.

% \textbf{Discussion}. Crucially, the stage-oriented planning operates exactly as intended in both scenarios. These failures highlight the inherent challenges LLMs face in strict semantic-code translation and holistic table comprehension. Such limitations could be mitigated in future work by integrating explicit metadata descriptions or incorporating a global schema-checking sub-stage.

\section{Discussion}

While our study has revealed the high potential of stage-oriented plan-then-execute table reasoning, certain limitations and emerging challenges warrant future exploration. 

Our proposed \textsc{PoTable} exhibits strong effectiveness and explainability in scenarios involving standard-structured tables and semantically precise objectives.
However, \textit{long-context} and \textit{domain-specific} challenges remain to be addressed.
Specifically, in long-context scenarios, analyzing multiple large tables (\textit{e.g.}, relational databases \cite{DB2AI,MMQA}) can easily exceed LLM context windows. 
It remains challenging to balance the effectiveness and efficiency when processing complex table relations and diversified content semantics. 
Additionally, when handling tasks with lengthy descriptions \cite{EHRCon}, it necessitates a comprehensive and precise understanding with advanced mechanisms like sub-task decomposition and fine-grained table retrieval.
Furthermore, for domain-specific scenarios (\textit{e.g.}, biology, finance), LLMs may hallucinate due to the lack of specialized knowledge or sparse attention to specific modalities (\textit{e.g.}, affiliated images, surrounding text). 
To address this issue, effective integration of external knowledge bases \cite{CKBQA,KGQA} and comprehension of multi-modal data \cite{TAT-QA,EVQA} are necessary in such domain-specific hybrid reasoning scenarios.

Recently, powerful reasoning LLMs (\textit{e.g.}, OpenAI-o1 \cite{o1}, DeepSeek-R1 \cite{R1}) have evolved the reasoning paradigm through extensive reinforcement learning that promotes the capabilities of internal CoT generation. 
Despite their promising performance, \textsc{PoTable} still maintains its advantage on table reasoning. 
Firstly, while these reasoning LLMs excel at implicit logical deduction, they generate free-form text with potential hallucinations \cite{o1-hallu} that are difficult to discover and verify in structured table reasoning. 
In contrast, \textsc{PoTable} produces fully executable Python programs with in-time verification, ensuring strict reproducibility and transparency. 
Secondly, employing such heavyweight reasoning LLMs for every single operation in a tabular workflow inevitably leads to overthinking \cite{overthinking}, while \textsc{PoTable} circumvents this by utilizing cost-effective LLMs under systematic guidance. 
Ultimately, \textsc{PoTable} is in fact \textit{orthogonal} to these advanced LLM backbones. 
Future improvements could selectively integrate advanced reasoning models solely for the complex planning phase, thereby enhancing task decomposition while preserving efficiency.

To address the above limitations, we will explore more
effective and efficient methods in the future to make our
study more adaptable and competitive in more extensive
and complicated table reasoning scenarios.

\section{Conclusion}

In this paper, we have proposed \textsc{PoTable}, a novel LLM-based framework that enables stage-oriented plan-then-execute reasoning on tabular data. 
By deploying well-defined analytical stages with explicit objectives, \textsc{PoTable} provides structured guidance for systematic, stage-by-stage reasoning. 
Within each stage, \textsc{PoTable} first tailors focused operation plans and then sequentially executes them through dynamic code generation, real-time running and feedback processing. 
Through the collaboration between an LLM and a Python interpreter, \textsc{PoTable} yields highly accurate, step-wise commented, and fully executable programs, closely mirroring the rigorous workflow of a professional tabular data analyst. 
Extensive experiments on two table reasoning benchmarks have demonstrated its superior performance over competitive baselines, firmly establishing its effectiveness, efficiency, and explainability.

\section*{Acknowledgments}
This work was supported by grants from the National Natural Science Foundation of China (62525606, 62502486), the Fundamental Research Funds for the Central Universities (WK2150110032) and the Provincial Natural Science Foundation of Anhui Province (2408085QF193).

\bibliographystyle{IEEEtran}
\bibliography{references}

% {\appendix[Proof of the Zonklar Equations]
% Use $\backslash${\tt{appendix}} if you have a single appendix:
% Do not use $\backslash${\tt{section}} anymore after $\backslash${\tt{appendix}}, only $\backslash${\tt{section*}}.
% If you have multiple appendixes use $\backslash${\tt{appendices}} then use $\backslash${\tt{section}} to start each appendix.
% You must declare a $\backslash${\tt{section}} before using any $\backslash${\tt{subsection}} or using $\backslash${\tt{label}} ($\backslash${\tt{appendices}} by itself
%  starts a section numbered zero.)}


% \input{bio/bio}

\end{document}